\documentstyle[prd,aps,epsfig,preprint,tighten]{revtex}

\def\cs{c_2}
\def\ro{{d_1}}

\def\si{\sigma}
\def\mm{M_{\tau}^2}

\def\Lms{\Lambda_{\overline{\rm{MS}}}}
\def\Lmss{\Lambda_{\overline{\rm{MS}}}^2}
\newcommand{\beg}{\begin{equation}}
\newcommand{\en}{\end{equation}}
\begin{document}
\thispagestyle{empty}
\title{Renormalization Scheme and Higher Loop Stability
in Hadronic $\tau$~Decay within Analytic Perturbation Theory }
\author{K.A. Milton$^1\thanks{e-mail: milton@mail.nhn.ou.edu}$,
I.L.~Solovtsov$^{1,2}\thanks{e-mail: solovtso@thsun1.jinr.ru}$,
O.P.~Solovtsova$^{1,2}$, V.I.~Yasnov$^{2,3}$}
\address{
\noindent
$^1$Department of Physics and Astronomy, University of
Oklahoma, Norman, OK 73019 USA \\
{}$^2$Bogoliubov Laboratory of Theoretical Physics, Joint
Institute for Nuclear Research, Dubna, Moscow Region, 141980, Russia \\
{}$^3$Department of Physics and Astronomy, University of Southern
California, Los Angeles, CA 90089-0484 USA
}
\maketitle

\begin{abstract}
We apply an analytic description to the inclusive
decay of the $\tau$ lepton. We argue that this method gives not only a
self-consistent description of the process both in the timelike region by
using the initial expression for $R_\tau$ and in the spacelike domain by
using the analytic properties of the hadronic correlator, but also leads to
the fact that theoretical uncertainties associated with unknown higher-loop
contributions and renormalization scheme dependence can be reduced
dramatically.
\end{abstract}

\section{ Introduction}

The ratio of hadronic to leptonic widths for the inclusive decay of the
$\tau$-lepton,
$R_{\tau}=\Gamma(\tau^{-}\to{hadrons}\;\nu_{\tau})/
\Gamma(\tau^{-}\to{\ell}\;{\bar{\nu}}_{\ell}\;{\nu}_{\tau})$, gives
important information about the QCD running coupling at relatively
small energy scales. The theoretical analysis of the hadronic decay of a
heavy lepton was performed in~\cite{Tsai} before the experimental
discovery of the $\tau$-lepton in 1975. Since then, the properties
of the $\tau$ have been
studied very intensively. Numerous publications are devoted to the QCD
description of the inclusive decay of the $\tau$-lepton and determination of
the QCD running coupling $\alpha_s$ at the $\tau$ mass scale. A
detailed consideration of this subject has been given in~\cite{BNP92}.
Recently, an updated QCD analysis has been performed by the
ALEPH~\cite{ALEPH98} and OPAL~\cite{OPAL99} collaborations, where
applications of different theoretical approaches to the $\tau$-decay have
been analyzed.

At present, the $R_{\tau}$-ratio is known experimentally to high accuracy,
$\sim0.5\%$. Nevertheless, the value of $\alpha_s$ extracted from
the data has a rather large error, in which
theoretical uncertainties are dominant. For example, the ALEPH
Collaboration result is $\alpha_s(M_\tau=1.777\,{\mbox{\rm GeV}})=
0.334\pm0.007_{\rm expt}\pm 0.021_{\rm theor}$~\cite{ALEPH98}. It should be
emphasized that nonperturbative terms, the values of which are not
well known, do not dominate these uncertainties, because their
contribution is rather small~\cite{BNP92,ALEPH98,OPAL99}. The main difficulty
is associated with the perturbative description.

The original theoretical
expression for the width $\Gamma(\tau^{-}\to{hadrons}\;
\nu_{\tau})$ involves integration over small values of timelike
momentum~\cite{Tsai}. The perturbative description with the standard running
coupling, which has unphysical singularities, becomes ill-defined in this
region and
some additional ansatz has to be applied to get a finite result for the
hadronic width. To this end, one usually  transforms to a
contour representation for $R_\tau$~\cite{Lam-Yan}, which allows one to give
meaning to the initial expression and, in principle, perform calculations in
the framework of perturbative QCD.  Assuming
the validity of this transformation it is possible
to present results in the form of a truncated power series with
$\alpha_s(M_\tau)$ as the expansion parameter~\cite{Braaten88,BNP92}. There
are also other approaches to evaluating the contour integral. The
Le~Diberder and Pich prescription~\cite{LP92} allows one to improve the
convergence properties of the approximate series and reduce the
renormalization scheme (RS) dependence of theoretical predictions. The
possibility of using different approaches in the perturbative description of
$\tau$-decay leads to an uncertainty in the value of $\alpha_s(M_\tau)$
extracted from the same experimental data. Moreover, any perturbative
description is based on this contour representation, {\it i.e.}, on the
possibility of converting the initial expression involving integration over
timelike momenta into a contour integral in the complex momentum plane.
To carry out this transition by using Cauchy's theorem requires certain
analytic properties of the hadronic correlator or of the corresponding Adler
function. However, the
required analytic properties are not automatically maintained in
perturbative QCD resummed by the
renormalization group. It is well known that at
the one-loop level the so-called ghost pole occurs in the invariant charge.
Higher-loop corrections do not solve this problem, but merely add some
unphysical branch points. The occurrence of incorrect analytic properties in
the conventional perturbative approximation makes it impossible to exploit
Cauchy's theorem in this manner and therefore prevents rewriting the initial
expression for $R_\tau$ in the form of a contour integral in the complex
momen\-tum-plane.

In this paper we will use the analytic approach proposed in~\cite{SS96-97}
(see also~\cite{SS99} for details).
Being inspired by  K\"all$\acute{{\rm e}}$n--Lehmann analyticity, which
is based on general principles of quantum field theory, this method ensures that
the running coupling possesses the correct analytic properties, leads to a
self-consistent definition of the effective charge in the timelike
region~\cite{MSS-timelike,newEM} (which cannot be a symmetrical reflection
of the spacelike one~\cite{st-MS99}), and provides equality between the
initial $R_{\tau}$-expression and
the corresponding contour representation~\cite{MSS97}. A
distinguishing feature of the analytic approach is the existence of a
universal infrared limiting value of the analytic running coupling at
$q^2=0$ which is independent of both the QCD scale parameter $\Lambda$\ and the
choice of renormalization scheme. This limiting value is defined by the general
structure of the Lagrangian and turns out to be stable with respect to
higher-loop corrections in contrast to the corresponding
quantity in conventional perturbation theory~(PT).
The higher-loop stability of the analytic perturbation theory (APT)
holds also for physical observables~\cite{MSS97,ee,Bj-MSS97,GLS-MSS98}.

However, it is not sufficient to study the stability with respect to
higher-loop corrections; one must also investigate the
stability with respect to choice of renormalization scheme. This is also
essential in order to estimate the uncertainty of the results
obtained. The theoretical ambiguity which is connected with higher-loop
corrections and with RS dependence becomes considerable at low energy
scales (see, e.g.,~\cite{ree}). The APT method, as an
invariant analytical version of
perturbative QCD~\cite{DV_TMF}, improves the situation and gives very
stable results over a wide range of renormalization schemes. This has been
demonstrated for the $e^+e^-$ annihilation ratio~\cite{ee} and for the
Bjorken~\cite{Bj-MSS97} and Gross--Llewellyn Smith~\cite{GLS-MSS98} deep
inelastic scattering sum rules.

The main aim of the paper is a study of the RS dependence which appears
in the description of the inclusive $\tau$ decay within the APT approach. We
will consider the $R_{\tau}$-ratio at the next-to-next-to-leading order
(NNLO) and the next-to-leading order (NLO) and compare results obtained with
those of standard perturbation theory.
\section{ QCD parametrization of $R_{\tau}$}

The ratio of hadronic to leptonic $\tau$-decay widths
can be written as
\beg
R_{\tau}=3\,S_{\rm EW}(|V_{ud}|^2+|V_{us}|^2)(1+\delta_{\rm QCD}),
\en
where $S_{\rm EW}=1.0194\pm 0.0040$~\cite{MarchS86} is the electroweak
factor, $|V_{ud}|= 0.9752 \pm 0.0007$ and $|V_{us}|=0.2218 \pm
0.0016$~\cite{PDG98} are the CKM matrix elements, and $\delta_{\rm QCD}$ is
the QCD correction (see~\cite{BNP92} for details).

We first introduce some definitions: ${\rm Im}\ \Pi\sim 1+r$ for the hadronic
correlator $\Pi (q^2)$\ and $ D\sim 1+d$ for the Adler function $ D(q^2)$.
Then for massless quarks
one can write $\delta_{\rm QCD}$\  as an integral over timelike
momentum $s$:
\beg
\label{to-tau}
\delta_{\rm QCD}=2\int_0^{\mm}\frac{ds}{\mm}{\left(1-\frac{s}{\mm}
\right)}^2\left(1+2\frac{s}{\mm}\right) r(s).
\en
Within the conventional perturbative approximation of $r(s)$ this integral
is ill-defined due to unphysical singularities of the running coupling lying
in the range of integration. The most useful trick to rescue the situation
is to appeal to analytic properties of the hadronic correlator $\Pi (q^2)$.
This opens up the possibility of exploiting Cauchy's theorem by rewriting
the integral in the form of a contour integral in  the complex $q^2$-plane
with the contour being a circle of radius $\mm$:
\beg
\label{contour} \delta_{\rm QCD}=\frac{1}{2\pi
i}\oint_{|z|=\mm}\frac{dz}{z}{\left(1-\frac{z}{\mm}\right)}^3
\left(1+\frac{z}{\mm}\right) d(z).
\en

Starting from the contour representation (\ref{contour}) the PT description can
be developed in the following two ways (see, e.g., \cite{Braaten96}). One is
Braaten's approach~\cite{Braaten88} in which the
quantity~(\ref{contour}) is represented in the form of truncated power
series with the expansion parameter $\alpha_s(M_\tau^2)$. The NNLO
representation for $\delta_{\rm QCD}$ is written as follows
\beg
\label{delta_Br}
\delta_{\rm QCD}^{\rm Br} = a_\tau \, + r_1\, a_\tau^2 + r_2\,a_\tau^3 \> ,
\en
where $a_\tau\equiv{\alpha_s(M_\tau^2)}/{\pi}$. The coefficients  $r_1$ and
$r_2$ in the $\overline{\rm{MS}}$ scheme with three active flavors are
$r_1=5.2023$ and $r_2=26.366$~\cite{BNP92,r35}. The running coupling
satisfies the renormalization group equation:
\beg
\mu^2\frac{da}{d\mu^2}=-\frac{b}{2}a^2(1+c_1a+\cs a^2) \, ,
\en
where $b$, $c_1$ and $c_2$ are the $\beta$-function coefficients. For three
active flavors $b=9/2$, $c_1=16/9$\ and $c_2^{\overline{\rm{MS}}}=3863/864$.

In the approach of Le~Diberder and Pich (LP)~\cite{LP92}, the PT expansion
is applied to the $d$-function\footnote{We use the definition $q^2<0$ in
the Euclidean region. We have made a few changes in notation from that given
in~\cite{MSS97}: now $a=\alpha_s/\pi$, and consequently $d_1$ and $d_2$
are what we called $d_2$ and $d_3$ previously.}
\beg
\label{d_expand}
d(q^2)=a(q^2)+\ro a^2(q^2)+d_2 a^3(q^2) \, ,
\en
where in the $\overline{\rm{MS}}$-scheme $d_{1}^{\overline{\rm{MS}}}=1.6398$
and $d_2^{\overline{\rm{MS}}}=6.3710$~\cite{r35} for three active quarks.
Substituting Eq.~(\ref{d_expand}) into Eq.~(\ref{contour}) leads to the 
following
expansion, which is not a power series in $a$,
\beg
\label{delta_LP}
\delta_{\rm QCD}^{\rm LP} = A^{(1)}(a)+ d_1 \,A^{(2)}(a)+d_2\,A^{(3)}(a)
\en
with
\beg
\label{A}
A^{(n)}(a)=\frac{1}{2\pi
i}\oint_{|z|=\mm}\frac{dz}{z}{\left(1-\frac{z}{\mm}\right)}^3
\left(1+\frac{z}{\mm}\right) a^{n}(z)\, .
\en

As noted above, transition to the contour representation~(\ref{contour})
requires certain analytic properties of the correlator, namely, that it must
be an analytic function in the complex $q^2$-plane with a cut along the
positive real axis.  The correlator parametrized by the PT running coupling
does not have this virtue \cite{s10,MSS97}. Moreover, the conventional
renormalization group method determines the running coupling in the spacelike
region, whereas the initial expression (\ref{to-tau}) for $R_{\tau}$
contains an integration over timelike momentum. Thus, we are in need of some
method of continuing the running coupling from the spacelike to the timelike
region that takes into account the proper analytic properties of the running
coupling~\cite{newEM}. Because of this failure of analyticity,
Eqs.~(\ref{to-tau}) and (\ref{contour}) are not equivalent in the framework
of PT~\cite{MSS97} and if one remains within PT, nothing can be said about
the errors introduced by this transition.

The analytic approach may eliminate these problems. To make our analysis
more transparent and to  demonstrate clearly the differences between the
consequences of the PT and APT methods, we restrict  our consideration here
to massless NNLO. The NNLO analysis can be performed in a more rigorous way
without model assumptions that allows us to avoid minor details and exhibit
the principal features of the APT approach. Thus, other effects, such as
nonperturbative terms, higher-loop corrections, and renormalon contributions
lie outside of the purpose of this paper. Note, however, that the NNLO
approximation is adequate to describe the actual physical situation because
numerically the corresponding terms give the principal contribution to the
$R_\tau$-ratio.

The function $d(q^2)$, which is analytic in the cut $q^2$-plane, can be
expressed in terms of the effective spectral function $\rho (\si)$, the
basic quantity in the APT method,
\beg
\label{drho}
d(q^2)=\frac{1}{\pi}\int^{\infty}_0
\frac{d\si}{\si - q^2}\, \rho(\si)\,.
\en
The connection between the QCD corrections to the $D$- and $R$-functions can be
written down in the form of the dispersion integral
\beg
d(q^2)=-q^2\int^{\infty}_{0} \frac{ds}{{(s-q^2)}^2}\,r(s)\, ,
\en
which is inverted by the following formula~\cite{Radyshkin82}
\beg
\label{r(s)}
r(s)=-\frac{1}{2\pi i}\int^{s+i\epsilon}_{s-i\epsilon}\frac{dz}{z}\,d(z)\, .
\en
Here, the contour lies in the region of analyticity of the $D$-function.
In terms of $\rho(\si)$ the function $r(s)$ defined for timelike
momenta can be expressed as follows~\cite{MSS-timelike}:
\beg
\label{rrho}
r(s)=\frac{1}{\pi}\int^{\infty}_{s}\frac{d\si}{\si} \rho(\si)\, .
\en
Eqs.~(\ref{drho}) and (\ref{rrho}) determine the QCD corrections $d(q^2)$,
which is defined in the Euclidean (spacelike)
region of momenta, and $r(s)$ defined for
the Minkowskian (timelike) argument, in
terms of the spectral function $\rho(\si)$.  For
$\delta_{\rm QCD}$, using Eq.~(\ref{to-tau}) or equivalently
Eq.~(\ref{contour}), in terms of $\rho(\si)$, we find\footnote{ To
distinguish APT and PT cases, we will use subscripts ``an'' and ``pt''.}
\beg
\label{danrho}
\delta_{\rm an}=\frac{1}{\pi}\int_0^{\infty}\frac{d\si}{\si}\rho (\si)-
\frac{1}{\pi}\int^{\mm}_0\frac{d\si}{\si}{\left(1-\frac{\si}{\mm}\right)}^3
\left(1+\frac{\si}{\mm}\right)\rho (\si) \, .
\en

In the APT approach, the spectral function is defined as the imaginary part
of the perturbative approximation to $d_{\rm pt} (q^2)$ on the physical cut:
\beg
\label{e.13}
\rho (\si)=\varrho_0 (\si)+d_1 \varrho_1 (\si)+d_2\varrho_2 (\si),
\en
where
\beg \label{e.14}
\varrho_n (\si)={\rm Im}[a^{n+1}(\si+i\epsilon)]\,.
\en
Substituting Eq.~(\ref{e.13}) into Eq.~(\ref{danrho}), we can rewrite
$\delta_{\rm an}$ in the form of the APT expansion
\beg
\label{delta_an}
\delta_{\rm an}= \delta^{(0)} \, + d_1 \,\delta^{(1)}  + d_2\,\delta^{(2)}  \, .
\en
Note that the APT representations of the $d$-function and the QCD correction
$\delta_{\rm QCD}$ are not in the form of power series.

The function $\varrho_0 (\si)$ in Eq.~(\ref{e.13}) defines the analytic
spacelike running coupling
\beg
a_{\rm an}(q^2)=\frac{1}{\pi}\int_0^\infty\frac{d\si}{\si-q^2}\,\varrho_0(\si).
\en
In the one-loop approximation it leads to \cite{SS96-97}
\beg
\label{one-loop}
a_{\rm an}(q^2)=a_{\rm pt}(q^2)\,+ \,{2\over b}
\frac{\Lambda^2}{\Lambda^2+q^2}\, .
\en
Unlike the one-loop PT running coupling, $a_{\rm
pt}(q^2)=2/b\ln{(-q^2/\Lambda^2)}\,$, the analytic running coupling
(\ref{one-loop}) has no unphysical ghost pole and, therefore, possesses the
correct analytic properties, arising from  K\"all\'en-Lehmann analyticity
that reflects the general principles of the theory. The non\-pertur\-bative
(non-logarithmic) term, which appears in the analytic running coupling, does
not change the ultraviolet limit of the theory and thus
the APT and the PT approaches
coincide with each other in the asymptotic region of high energies.

Thus, the APT approach provides a self-consistent description of the
hadronic $\tau$ decay. This description can be equivalently phrased
either on the
basis of the original expression (\ref{to-tau}), which involves the
Minkowskian quantity $r(s)$, or on the contour
representation~(\ref{contour}), which involves
the Euclidean quantity $d(q^2)$.

An important feature of the APT approach is the fact that $d_{\rm an} (q^2)$
and $a_{\rm an}(q^2)$  have a universal limit at the point $q^2=0$. This
limiting value, generally, is independent of both the scale parameter
$\Lambda$ and the order of the loop expansion being considered. Because
$d_{\rm an}(0)$\ and $a_{\rm an}(0)$\  are equal to the reciprocal of the
first coefficient of the QCD $\beta$-function, they are also RS invariant
(we consider only gauge- and mass-independent RSs). The existence of this
fixed point plays a decisive role in the improved convergence
properties relative to PT and in the
very weak RS dependence of our results.

To find the analytic function $d(q^2)$ involved in Eq.~(\ref{contour}),
we solve the transcendental equation for the running coupling
\beg
\label{a-RS}
\frac{b}{2}\ln\left(\frac{-q^2}{\Lmss}\right)
-i\pi\frac{b}{2}=d_1^{\overline{\rm{MS}}}- \ro+\frac{1}{a}+c_1\ln
\left(\frac{b}{2c_1}\right)+F^{(l)}(a),
\en
where at the NLO
\beg
F^{(2)}(a)= c_1\ln\left(\frac{c_1a}{1+c_1a}\right),
\en
and at the NNLO
\beg
F^{(3)}(a)=F^{(2)}(a)+ c_2\int^a_0\frac{dx}{(1+c_1x)(1+c_1x+c_2x^2)}\, ,
\en
on the physical cut lying along the positive real axis in the complex
$q^2$-plane and then use Eqs.~(\ref{e.13}), (\ref{e.14}) and (\ref{drho}).
Eq.~(\ref{a-RS}) holds in any MS-like renormalization scheme and allows us
to normalize the results obtained by using the scale parameter
$\Lambda_{\overline{\rm MS}}$. Having found $\Lms$, we can study how
$\delta_{\rm an}$ varies with a change of renormalization scheme. To do that
one has to select parameters which determine the RS. The function $d(q^2)$
in Eq.~(\ref{contour}) is parametrized by a set of RS-dependent parameters.
There are RS invariant combinations which constrain these parameters~\cite{r42}.
At the NNLO there are two RS-invariant quantities; the first of them
expresses an energy dependence, the second is just a number
\beg\label{rgi}
\omega_2=\cs+d_2-c_1\ro-\ro^2\, ,
\en
which in our case equals 5.2378. Here, $c_1$ is RS invariant and we can
choose $\ro$ and $\cs$ as independent variables, which define some RS.

There are no fundamental principles upon which one can choose one or another
preferable RS. Nevertheless, a natural way of studying the RS dependence is
to supplement results in a certain scheme with an estimate of the variability
of the predictions over a range of {\it a priori\/} acceptable schemes
specified by some criterion.
In~\cite{r59} it was proposed to consider the class of `natural' RSs,
which obey the condition
\beg
\label{domain}
|\cs | +|d_2|+c_1|\ro|+\ro^2\le C |\omega_2|\, .
\en
This inequality is called the ``cancellation index criterion" which means
that the degree of cancellation in the second RS invariant (\ref{rgi}) should
not be too large. To define a boundary of `acceptable' schemes which is
defined by the value of the cancellation index $C$, we will require
no more cancellation than that which occurs in the scheme obeying
 the principle of minimal sensitivity (PMS)~\cite{r}, which
leads to $C\simeq2$.

\section{APT: convergence properties and RS dependence}
For various physical quantities, the APT approach allows one to construct
a series that has improved convergence properties as compared to a
perturbative expansion. To demonstrate this fact for the hadronic
$\tau$-decay, we compare the convergence properties
of the PT expansions (\ref{delta_Br}) and (\ref{delta_LP}) on the one hand,
and the APT approach given by Eq.~(\ref{delta_an}) on the other hand. For
our calculation we take as input the TAU'98 conference value:
$R_{\tau}=3.642 \pm 0.019$~\cite{TAU98}, which is consistent with the PDG'98
fit $R_{\tau}=3.642 \pm 0.024$~\cite{PDG98}. In Table~{\ref{T1} we present
NNLO results obtained by the methods mentioned above  for the central
experimental value in the $\overline{\rm{MS}}$ scheme. The
relative contributions of higher-order terms
depends on the method which is applied. The convergence properties
of the APT expansion seem to be much improved compared to those of
the PT expansions.

The values of the scale parameter $\Lambda_{\overline{\rm{MS}}}$ and the
coupling $\alpha_s(M_\tau^2)$ obtained from above PT and APT expansions are
noticeably different from each other. The corresponding numerical estimations
are given in Table~\ref{Tab1}, in which, in order to clarify the situation
concerning higher-loop stability of different expansions, we also present the
NLO result. This table demonstrates that the theoretical ambiguity, which
associated with different versions of the perturbative description, leads to
a rather large uncertainty, $\alpha_{\rm PT~(Br)}^{\rm NNLO}-\alpha_{\rm
PT~(LP)}^{\rm NNLO}=0.012$. At the same time the experimental error is
$\Delta\alpha_{\rm expt}=0.007$--$0.009$~\cite{ALEPH98,OPAL99}. The
distinction between NLO and NLLO running coupling values is $12\%$ for
PT~(Br) and $5\%$ for PT~(LP) approaches, while for the APT approach it is
less than $0.5\%$.

The non-logarithmic terms, which ensure the correct analytic properties and
allow a self-consistent description of $\tau$ decay, turn out to
be very important for the numerical analysis and influence in an essential
way the
value of $\Lambda$ parameter extracted from the data. Indeed, at the
one-loop level one can write a simple relation: $\delta_{\rm
an}(\Lambda)\simeq \delta_{\rm pt}^{\rm LP}
(\Lambda)-(2/b){\Lambda^2}/{M_{\tau}^2}$. The second term, which is
`invisible' in the perturbative expansion, turns out to be numerically
important~\cite{OlSol96} (see the detailed discussion in~\cite{newEM}). Note
also that there is a difference between the shapes of the analytic and
perturbative running couplings, for example, $\alpha_{\rm
an}(\Lambda=907\,{\mbox{\rm MeV}}) =0.403$, while at the same
scale, the value of
the perturbative coupling much larger, $\alpha_{\rm pt}(\Lambda=907\,
{\mbox{\rm MeV}})=0.796$. Here the question may arise, how is
the large APT value of
$\Lambda$ consistent with high energy experimental data? We have
estimated the ratio of hadronic to leptonic Z-decay widths, $R_{\rm Z}$,
using the above value of $\Lambda_{\rm an}$
and the matching procedure proposed in~\cite{newEM}. We obtained the
value $R_{\rm Z}=20.82$, which lies within the range of experimental
errors; for example, the PDG'98 average is
 ${\rm R}_{\rm Z}=20.77\pm0.07$~\cite{PDG98}. This fact can be
understood if one takes into account that there are differences between
the shapes of the analytic and perturbative running couplings and also in the
terms of the corresponding series.

We found that the value of $\delta_{\rm an}$ depends so slightly on $\Lms$
that a $0.9\%$ error in $R_{\tau}$ gives $18\%$\ error in the value of
$\Lms$. (This is the reason why the errors in the values of
$\Lambda_{\overline{\rm MS}}$ and $\alpha(\mm)$ given by APT are larger than
those in PT.)
We illustrate this feature in Table~\ref{Tab2}. According to the table,
when we change $\mm/\Lambda^2$ from 2.0 to 6.5 (corresponding to a variation
of $\Lambda$ from 1.256~GeV to 0.697~GeV), $\delta_{\rm an}$ is only altered
by about $20\%$. The sensitivity to $\Lms$\ increases as $\mm/\Lambda^2$
gets smaller.

Consider now the RS dependence of the APT result and compare it with
the perturbative LP approach,\footnote{The LP approach is often called the
contour-improved fixed-order PT (CIPT)~\cite{TAU98}.} which of the two PT
schemes is more
preferable from the viewpoint of sensitivity to RS dependence.
In the framework of the PT, the RS dependence of $\delta_{\rm QCD}$
has been discussed  in detail in~\cite{r}.

In the $\overline{\rm{MS}}$ scheme we adopt $\delta_{\rm an,pt}=0.1906$
and consider some RS belonging to the domain described above
[see Eq.~(\ref{domain})]. Take two schemes, $A$ and $B$,
located at the two lower corners of
the boundary of the domain (see Fig.~\ref{fig1}), {\it i.e.}, they have
the same cancellation index as does the PMS scheme, with
$A=\left(-1.6183,0\right)$ and $B=\left( 0.9575,0\right)$, where the first
coordinate is $\ro$\ and the second is $\cs$.
Then for the PT case in NNLO we get $\delta_{\rm pt}(A)=0.2025$
and $\delta_{\rm pt}(B)=0.1911$.
Therefore, even for this sufficiently narrow class of RS the perturbative
approach gave a $6\%$ deviation in $\delta_{\rm QCD}$ that corresponds to
a RS uncertainty for the running coupling value in the
$\overline{\rm{MS}}$-scheme of $\Delta\alpha^{\rm RS}_{\rm pt}=0.0153$.
The difference between APT results is much smaller: $\delta_{\rm an}
(A)=0.1890$\ and $\delta_{\rm an} (B)=0.1905$, and we have only $0.8\%$\
deviation, which corresponds in the $\overline{\rm{MS}}$-scheme
to a RS uncertainty of
$\Delta\alpha^{\rm RS}_{\rm an}=0.0035$. The similar RS
stability holds also at the two-loop level: one has a $5\%$ deviation in the
PT case  and only a $0.4\%$ for the APT. We display our NNLO results
in the form of a contour plot, in Fig.~\ref{fig1}.

It is worthwhile to analyze some schemes lying outside the domain considered
above with the relatively small value of the cancellation index
$C\simeq2$. Among them there is, for instance, the commonly used
$\overline{\rm{MS}}$ scheme which does not belong this domain. In~\cite{r}
it was shown that the so-called $V$ scheme~\cite{r61} lies
very far from the domain described above and gives so a large value of
$\delta_{\rm pt}$ that it cannot be used at this low energy. For the
$V$ scheme we have $\ro=-0.109$\ and $\cs=26.200$. The three-loop
perturbative result is $\delta_{\rm pt}(V)=0.3060$ that corresponds to about
a $61\%$\ deviation from the $\overline{\rm{MS}}$ scheme. On the other hand,
if we turn to APT we have $\delta_{\rm an}=0.1902$, i.e., only about a
$0.2\%$ deviation from the $\overline{\rm{MS}}$ scheme. So the $V$ scheme is
still useful at this energy in APT.

In PT at high energies the weak RS dependence is a consequence of the
small value of the coupling constant. At lower energies the uncertainty
increases. In APT, at high energies, the situation is the same.
However, at low energies the theory has a universal RS-invariant infrared
limiting value $d_{\rm an}(0)$, which restricts the RS ambiguity over a very
wide range of momentum.
Another way to illustrate the remarkable stability of APT is to calculate
the spectral functions $\varrho_n (\si)$ given by Eq.~(\ref{e.14}); one sees
that $\varrho_1(\si)$ is much smaller than $\varrho_0 (\si)$ over the whole
spectral region. The same statement is true for the relationship between
$\varrho_1 (\si)$ and $\varrho_2 (\si)$. This monotonically decreasing
behavior reduces the RS dependence strongly, since the perturbative
coefficients $d_1$\ and $d_2$\ in expression  (\ref{e.13}) for $\rho(\si)$
are multiplied by these functions. For the $\overline{\rm MS}$ scheme, this
situation is demonstrated in Fig.~\ref{rho-012}.

\section{Conclusion}
We have considered inclusive $\tau$-decay in three-loop order within
analytic perturbation theory concentrating on the analysis of theoretical
uncertainties coming from the perturbative short distance part of the QCD
correction to the $R_\tau$-ratio, which defines the principal 
contribution to this
physical quantity. For the low energy $\tau$-mass scale, the main 
source of theoretical uncertainties  results from the inevitable 
truncation of the perturbative series, which leads to the essential RS 
dependence and higher loop sensitivity of the theoretical predictions. 
In order to resolve this problem within the conventional perturbative 
approach it is possible to try, in principle, to compute higher 
loop contributions. However, even if this were to 
be done, one has to keep in mind 
that from the rigorous point of view
it will hardly be sufficient because the series is
asymptotic, and, in any finite order, the analytic properties of
the hadronic correlator, which arise from general principles of the theory, are
violated. Thus, to resolve this problem one has to use a modification of the
perturbative expansion at low energy scales.

Here, we have applied the analytic approach which is not inconsistent with the
general principles of quantum field theory and which opens up the possibility of
reducing the theoretical uncertainties associated with short
distance contributions mentioned above. 
Let us summarize the important features of this
method: (i) the method maintains the correct analytic properties and leads
to a self-consistent definition of the procedure of analytic continuation
from the spacelike to the timelike region; (ii) the APT approach has much
improved convergence properties and turns out to be stable with respect to
higher-loop corrections; (iii) the RS dependence of the results obtained is
reduced drastically. For example, the $V$ scheme, which gives a very large
discrepancy in standard perturbation theory, can be used in analytic
perturbation theory without any difficulty and the APT predictions are
practically RS independent over a wide region of RS parameters.

The nonperturbative power corrections coming from the operator product
expansion (in this connection see a discussion in~\cite{Grunberg97,DMW96}),
renormalon and other effects are beyond the scope of our present consideration.
Note, however, that the process of enforcing analyticity modifies the
perturbative contributions by incorporating some non\-per\-turbative terms.
The form of the APT running coupling and the non-power structure of the APT
expansion are essentially different from the PT ones. Numerically, this
difference becomes very important in the region less than a few GeV and in
order to get the same physical quantity the contribution of power
corrections should also be changed.

The value of $\Lambda_{\rm APT}$ is very sensitive to the experimental value
of $R_{\tau}$. For example, as has been demonstrated in \cite{MSS97} the
use of the value of $R_\tau$ obtained by the CLEO
collaboration~\cite{CLEO95} gives a value of the
scale parameter some 30\% smaller than that found here.
Note also that the renormalon contribution influences the
value of $\Lambda$ extracted from the $\tau$ data (see \cite{Beneke99} for
a review). Within the usual approach, renormalons reduce the
value of $\alpha_s(M_\tau^2)$ by 
about $15\%$. A similar situation holds also in APT and for the
nonperturbative $a$-expansion approach~\cite{JRS97}, which allows one, as
in APT, to maintain the required analyticity~\cite{s10}. These two
analytic approaches often lead to rather similar consequences. For example,
they allow one to get a good description of experimental data
corresponding to the Euclidean and Minkowskian characteristics of the process
of $e^+e^-$ annihilation into hadrons down to the lowest energy
scale~\cite{ee,VPT_D}.

Pure massless APT analysis, which has been performed here, leads to an unusually
large value of the QCD scale parameter $\Lambda$ as compared to the
conventional PT value. This is connected with the presence of
nonperturbative contributions that appear in the APT method which have a
negative relative sign. The effects mentioned above 
can change the value of the scale
parameter extracted from the $\tau$ data. However, this fact is not relevant
for the essential conclusion
which we have claimed in this paper, that the APT method provides
predictions which are stable with 
respect to the choice of renormalization scheme and to the inclusion of
higher loop corrections.
Thus, the analytic approach discussed here is not in conflict with the
general principles of the theory and allows one to reduce the uncertainties
of theoretical predictions drastically.

\section*{Acknowledgements}

The authors would like to thank D.V.~Shirkov for interest in this work.
Partial support of the work by the US National Science Foundation, grant
PHY-9600421, and by the US Department of Energy, grant DE-FG-03-98ER41066,
and by the RFBR, grants 99-01-00091, 99-02-17727 is gratefully acknowledged.
The work of ILS and OPS is also supported in part by the University of
Oklahoma, through its College of Arts and Science, 
the Vice President for Research, and the Department of Physics and Astronomy.

\begin{figure}[tbp]
\centerline{\epsfig{file=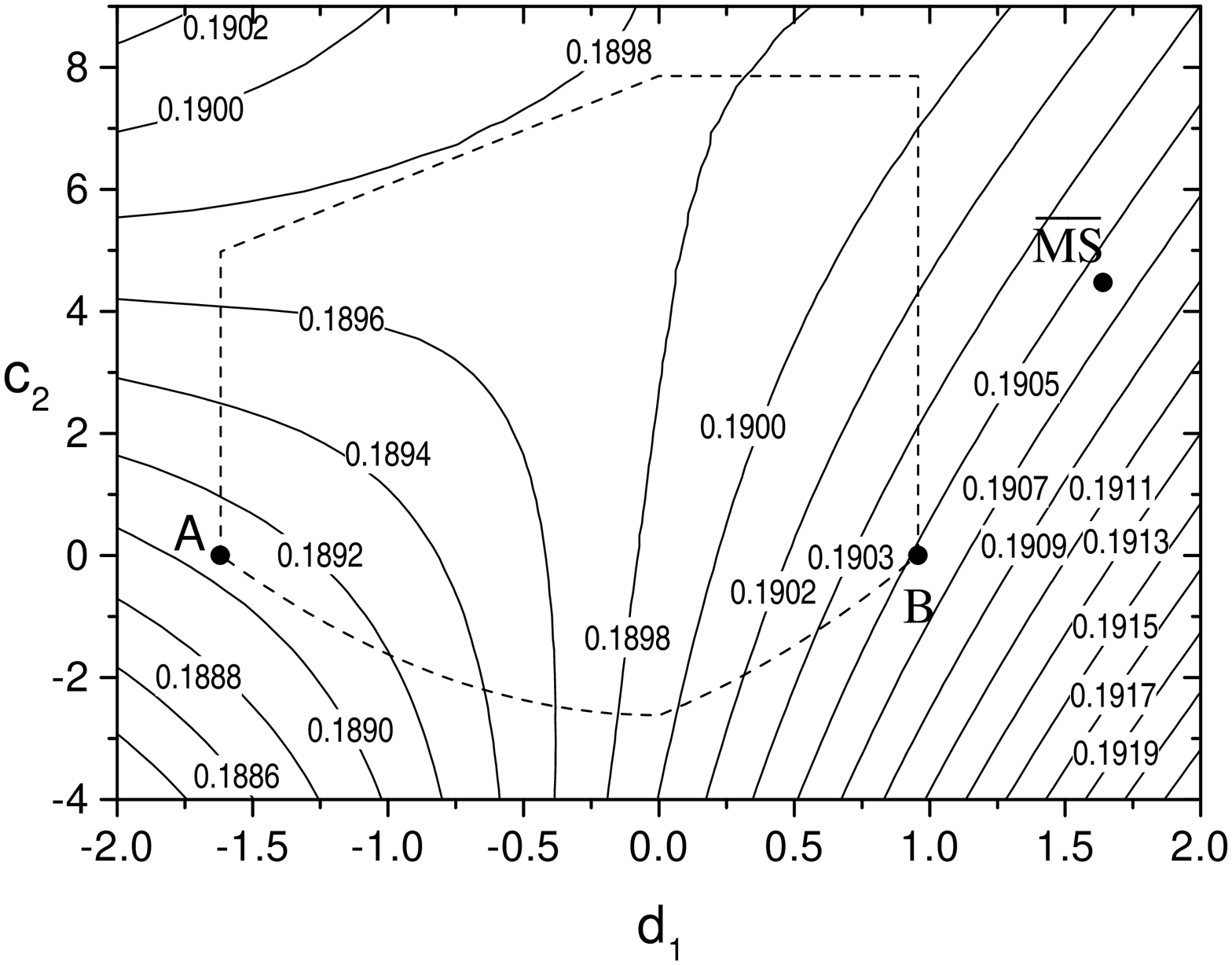,width=16.0cm}}
         \vspace{5mm}
         \caption{Contour plot of values of $\delta_{\rm an}$ at
     the three-loop order as a function of RS parameters $d_1$\ and
     $c_2$. The dashed line indicates the boundary of the domain,
     defined by Eq.~(\protect\ref{domain}) with $C=2$, the heavy
     points are
     the positions of the $A$, $B$ and ${\overline{\rm MS}}$ schemes.}
    \label{fig1}
\end{figure}


\begin{figure}[tbp]
\centerline{\epsfig{file=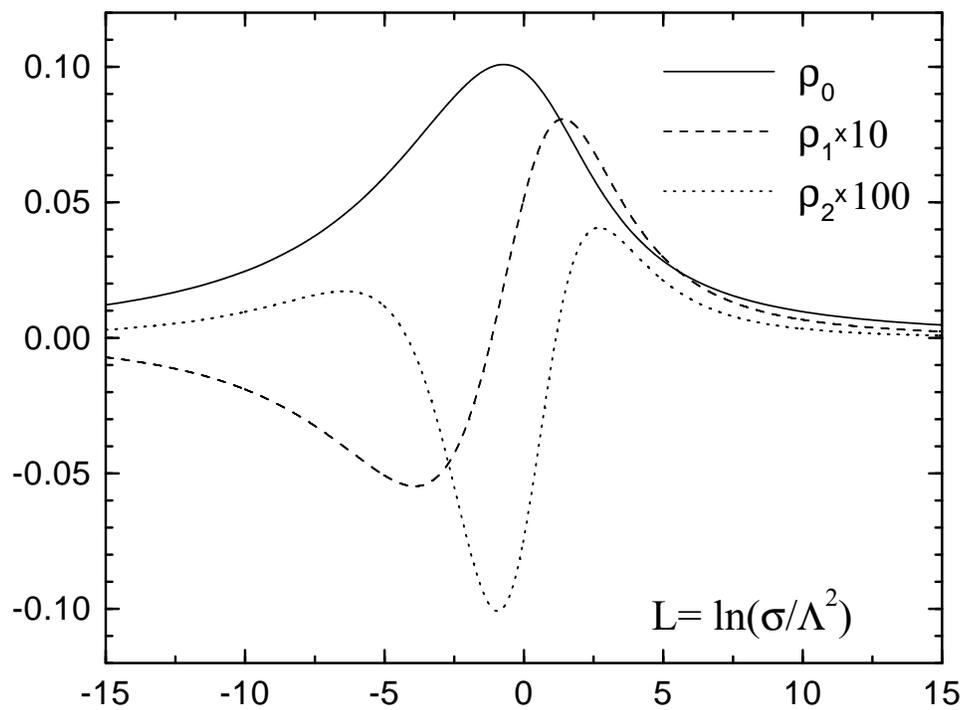,width=16.0cm}}
         \vspace{5mm}
         \caption{The spectral densities $\varrho_0$, $\varrho_1\cdot 10$
     and $\varrho_1\cdot 10^2$ vs. $L=\ln(\sigma/\Lambda^2)$ in
     the $\overline{\rm MS}$ scheme. }
    \label{rho-012}
\end{figure}

%
\begin{table}
\caption{Successive loop contributions to the PT and APT expansions for
$R_{\tau}/[3\,S_{\rm EW}(|V_{ud}|^2+|V_{us}|^2)]$.}

\hphantom{}
\label{T1}

\begin{tabular}{lc}
~~Method of description& Expansion terms   \\ [0.1cm] \hline
~~~~~PT~(Br)\cite{Braaten88}
&$1~+~\delta_{\rm pt}^{\rm Br}~=~1~+~0.104~+~0.056~+~0.030$
   \\
~~~~~PT~(LP)\cite{LP92}
&$1~+~\delta_{\rm pt}^{\rm LP}~=~1~+~0.148~+~0.030~+~0.012$
   \\
~~~~~APT~\cite{SS96-97}
& $1~+~\delta_{\rm an}~=~1~+~0.167~+~0.021~+~0.002$
\end{tabular}
\end{table}

\begin{table}[tbh]
\caption{QCD parameters extracted from
$R_{\tau}=3.642 \pm 0.019$~\protect\cite{TAU98}
in the $\overline{\rm{MS}}$ scheme.}

\hphantom{}

\begin{tabular}{clcc}
Approximation& Method &$\Lambda_{\overline{\rm MS}}$~(MeV)
&$\alpha(\mm)$\\ [0.1cm] \hline
NNLO &PT~(Br)  & $366\pm 14$ &$0.328\pm 0.007$   \\
      & PT~(LP) & $391\pm 16$ &$0.340\pm 0.008$  \\
      & APT & $ 907\pm 94$ &$0.403\pm 0.015$ \\
\hline
NLO  & PT~(Br)     & $492\pm 17$ &$0.371\pm 0.009$   \\
 &   PT~(LP)       & $465\pm 19$ &$0.358\pm 0.009$    \\
 & APT            & $954\pm 90$ &$0.404\pm 0.014$\\
\end{tabular}
\label{Tab1}
\end{table}


\begin{table}
\caption{NLO and NNLO predictions for $\delta_{\rm an}$ in the
$\overline{\rm{MS}}$\ scheme. }

\hphantom{}

\begin{tabular}{cccccc}
$\mm/\Lambda^2$ &$\delta_{\rm an}^{{\rm NLO}}$ &
$\delta_{\rm an}^{{\rm NNLO}}$ & $\mm/\Lambda^2$
&$\delta_{\rm an}^{{\rm NLO}}$ & $\delta_{\rm an}^{{\rm NNLO}}$ \\ [0.1cm]
\hline
2.0&0.2090&0.2106&4.5&0.1820&0.1857 \\
2.5&0.2016&0.2039&5.0&0.1785&0.1824 \\
3.0&0.1955&0.1983&5.5&0.1753&0.1795 \\
3.5&0.1904&0.1935&6.0&0.1724&0.1767 \\
4.0&0.1859&0.1894&6.5&0.1698&0.1743 \\
\end{tabular}
\label{Tab2}
\end{table}

\end{document}